\documentclass[12pt]{article}
\usepackage[dvips]{epsfig}
\newcommand{\PY}{{PYTHIA}}
\newcommand{\CO}{{CompHEP}}
\newcommand{\noi}{\noindent}
\newcommand{\eeinto}   {$e^+e^- \longrightarrow \:$}
\newcommand{\geinto}   {$\gamma e \longrightarrow \:$}
\newcommand{\into}   {$\longrightarrow \:$}
\newcommand{\entb}{$e \nu t b \: $}
\newcommand{\gentb}{$\gamma e \longrightarrow  \nu \bar{t} b \: $}
\newcommand{\hww}{$HWW \: $}
\newcommand{\nbbw}{$\nu \,b\, \bar{b}\, W\: $}
\newcommand{\wtb}{$W t b \: $}
\newcommand{\mt}{$m_{t} \: $}
\newcommand{\mh}{$M_{H} \: $}
\newcommand{\ee}{$e^+e^-\: $}
\newcommand{\vtb}{$|V_{tb}| \: $}
\newcommand{\toto}{$t\,\bar{t}\: $}
\newcommand{\tata}{$\tau^+\,\tau^-\: $}
\newcommand{\bb}{$b\,\bar{b}\:\; $}
\newcommand{\cc}{$c\,\bar{c}\:\; $}
\newcommand{\glgl}{$g\,g\: $}
\newcommand{\mm}{$\mu^+\mu^-\:$}
\newcommand{\qq}{$q\,\bar{q}\:\; $}
\newcommand{\nn}{$\nu\bar{\nu}\: $}
\newcommand{\gaga}{$\gamma\gamma\: $}
\newcommand{\ff}{$f\bar{f}\: $}
\newcommand{\lele}{$l\bar{l}\: $}
\newcommand{\ww}{$W^+W^- \: $}

\newcommand{\zz}{$ZZ \: $}
\newcommand{\znull}{$Z^0 \: $}
\newcommand{\hnull}{$H^0 \: $}
\newcommand{\gae}{$\gamma e\: $}
\newcommand{\nene}{$\nu_e\bar{\nu_e}\: $}
\newcommand{\SQRTS}{$\sqrt{s}\:$}
\newcommand{\SQRTSGE}{$\sqrt{s_{\gamma e}}\:$}
\newcommand{\SQRTSEE}{$\sqrt{s_{e^+e^-}}\:$}
\newcommand{\less}{\stackrel{ <}{\sim}}
\newcommand{\gess}{\stackrel{ >}{\sim}}
\newcommand{\Ptmiss}{$p_\perp^{miss}\:$}
\newcommand{\Ptb}{$p_{\perp}^b\:$}
\newcommand{\Ptbb}{$p_{\perp}^{b\bar{b}}\:$}
\newcommand{\Pt}{$p_\perp\:$}
\newcommand{\Ptt}{$p_{\perp}^t\:$}
\newcommand{\Pttot}{$p_{\perp}^{tot}\:$}
\def\PL #1 #2 #3 {Phys. Lett. {\bf#1}              (#3)  #2}
\def\MPL #1 #2 #3 {Mod. Phys. Lett. {\bf#1}        (#3)  #2}
\def\NP #1 #2 #3 {Nucl. Phys. {\bf#1}              (#3)  #2}
\def\PR #1 #2 #3 {Phys. Rev. {\bf#1}               (#3)  #2}
\def\PP #1 #2 #3 {Phys. Rep. {\bf#1}               (#3)  #2}
\def\PRL #1 #2 #3 {Phys. Rev. Lett. {\bf#1}        (#3)  #2}
\def\CPC #1 #2 #3 {Comp. Phys. Commun. {\bf#1}     (#3)  #2}
\def\ANN #1 #2 #3 {Annals of Phys. {\bf#1}         (#3)  #2}
\def\APP #1 #2 #3 {Acta Phys. Pol. {\bf#1}         (#3)  #2}
\def\ZP  #1 #2 #3 {Z. Phys. {\bf#1}                (#3)  #2}
\def\NIM  #1 #2 #3 {Nucl. Instr. and Meth. {\bf#1} (#3)  #2}
 
\newcommand{\BS}{\bigskip}
\newcommand{\VS}{\,\,\,\,}
\newcommand{\SECTION}[1]{\BS\begin{center}{\large\section{\bf #1}}\end{center}}
\newcommand{\SUBSECTION}[1]{\BS{\large\subsection{\bf #1}}}
\newcommand{\SUBSUB}[1]{\BS{\large\subsubsection{#1}}}

\begin{document}

\pagestyle{empty}

\noi DESY 97-227

\BS\BS

\noi November 1997
\section*{
\vspace{4cm}
\begin{center}
\LARGE{\bf
 Higgs Physics at a 300 - 500 GeV\\ $e^+e^-$ Linear Collider
          }\\
\end{center}
}

\vspace{2.5cm}
\large
\begin{center}
H. J\"urgen\, Schreiber${}^*$\\
\bigskip \bigskip  

DESY- Zeuthen, Germany \\
\end{center}
\vfill
{$*$ Contribution  to the Proceedings of the Internat. Europhysics Conf.
on High Energy Physics,
~16-26 August 1997, Jerusalem, Israel}

\newpage
\pagestyle{plain}
\pagenumbering{arabic}
\section*{Abstract}

We summarize discovery potentials for the Standard Model Higgs boson
produced in $e^+e^-$ collisions and measurements of its detailed
properties. Prospects for Higgs boson detection within the Minimal
Supersymmetric Standard Model are also discussed.
\BS\BS\BS


\large
\section{Introduction}

Electroweak breaking due to the Higgs mechanism \cite{Higgs} implies the
existence of at least one new particle, the Higgs boson. Its discovery
is the most important missing link for the formulation of electroweak
interactions. LEP has established a lower bound of the Higgs mass $M_H
\gess$ 77 GeV  \cite{lowM}. High precision data, interpreted within the
Standard Model (SM), favour a Higgs boson with a mass somewhere
between 100 and 180 GeV.

Future $e^+e^-$ linear colliders are the ideal machines
for a straightforward
Higgs boson discovery, its verification and precision measurements
of the Higgs sector properties.

Recent simulation studies \cite{our} involve, thanks to the effort
of several groups,\\
i) the full matrix elements for 4-fermion final states,
ii) all Higgs decay modes (with SM decay fractions $\gess $ 1\%),
iii) initial state QED and beamstrahlung (TESLA design), 
iv) a detector response \cite{CDR} and
v) all important background expected to contribute.

\section{SM Higgs discovery potentials}

The Higgs boson can be produced by the Higgsstrahlung process $e^+e^-$
$\rightarrow Z^* \rightarrow ZH^o$  (1), or by the fusion of WW and
ZZ bosons,   $e^+e^- \rightarrow \nu \bar{\nu} H^o$ (2) and 
$e^+e^- \rightarrow e^+e^- H^0$ (3), respectively,
or by 
  radiation off top quarks, $e^+e^- \rightarrow t \bar{t} H^o$, with
  however a very small cross section.
At e.g.
$\sqrt{s}$ = 360 GeV
and $M_H$ = 140 GeV, the Higgsstrahlung process is about four times
more important than the fusion reactions. Reaction (1) 
admits two strategies for the Higgs search: i) calculation of the recoil mass
against the $Z \rightarrow e^+e^- / \mu^+\mu^-, M^2_{rec} =$
$ s - 2\sqrt{s} (E_{l^+} + E_{l^-}) + M^2_Z$, which is independent of
assumptions about Higgs decay modes, and ii) 
the direct reconstruction of the invariant mass of the Higgs decay
products.

In order to achieve the best experimental resolution, energy-momentum
as well as $M(l^+l^-) = M_Z$ constraints have been imposed when
appropriate, and to make a signal-to-background analysis as meaningful
as possible a consistent evaluation of the signal
and all expected background
rates has been made \cite{our}.

The leptonic channel, $e^+e^- \rightarrow ZH^o \rightarrow $
$(e^+e^- / \mu^+\mu^-) (b\bar{b})$, allows Higgs detection either in
the mass recoiling against the Z or in the hadronic two b-quark jet
mass. Typically, an integrated luminosity of $\sim 10 fb^{-1}$ is
needed  to observe  the Higgs boson with a significance
$S/\sqrt{B} > 5$ after application of appropriate selection
procedures \cite{our}.

The tauon channel,  $e^+e^- \rightarrow ZH^o \rightarrow (q\bar{q})$
$(\tau^+\tau^-)$, requires some more refined selection procedures
\cite{our, Hil} due to missing neutrinos from $\tau$ decays.
Hence, the accumulated luminosity to observe a clear $H^o$ signal
in the recoil resp. $\tau^+\tau^-$  mass should be between 50 to
80 $fb^{-1}$. Energy-momentum constraints are required to determine
the original $\tau$ and jet energies by a fit; without that no $H^o$
signal would be visible.

The missing energy channels, $e^+e^- \rightarrow ZH^o \rightarrow$
$(\nu\bar{\nu})(b\bar{b})$ and $e^+e^- \rightarrow \nu\bar{\nu} H^o
\rightarrow \nu\bar{\nu} (b\bar{b})$ are very important due to their
large discovery potential for the Higgs boson. It has been
demonstrated \cite{our,Janot} that with $\sim 1 fb^{-1}$ of integrated
luminosity a convincing signal over some small remaining background in
the doubly-tagged b-jet mass should be obtained (Fig.1).
\begin{figure*}[h!t]
\begin{center}
\mbox{\epsfxsize=17cm\epsfysize=17.5cm\epsffile{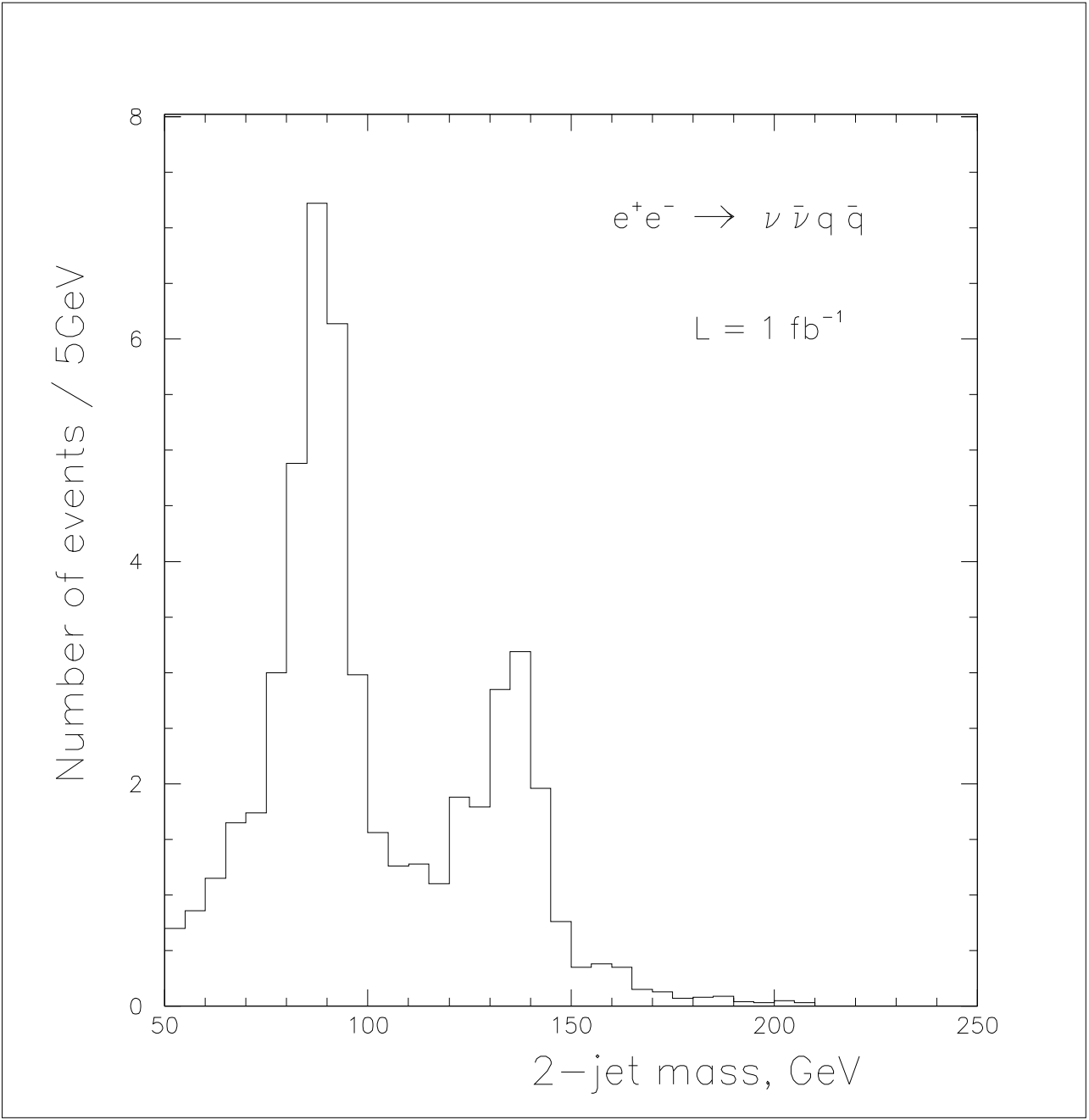}}
\end{center}
\caption[ ]{\sl 2-jet  mass distributions of the
reaction \ee \into \nn\qq,
~for an integrated luminosity of 1 fb$^{-1}$ at $\sqrt{s} =$ 360 GeV.
Background contributions are
included.}
\label{fig:neutrinos}
\end{figure*}
Thus, few days
of running a 300 GeV $e^+e^-$ linear collider at its nominal
luminosity would suffice to
 discover the SM Higgs in the intermediate mass range.

With increasing accumulated luminosity other $H^o$ decay modes like
$WW^{(*)}$ and $c\bar{c}$ + gg would be measurable, allowing to
 verify the Higgs
interpretation; 50 to 100 $fb^{-1}$ are needed for
significant signals. As an example, Fig.2 shows the signal from
 selected $H^o \rightarrow
c\bar{c}$ + gg decays in presence of a huge background
in the 2-jet missing energy event topology.
\begin{figure*}[h!t]
\begin{center}
\mbox{\epsfxsize=17cm\epsfysize=17.5cm\epsffile{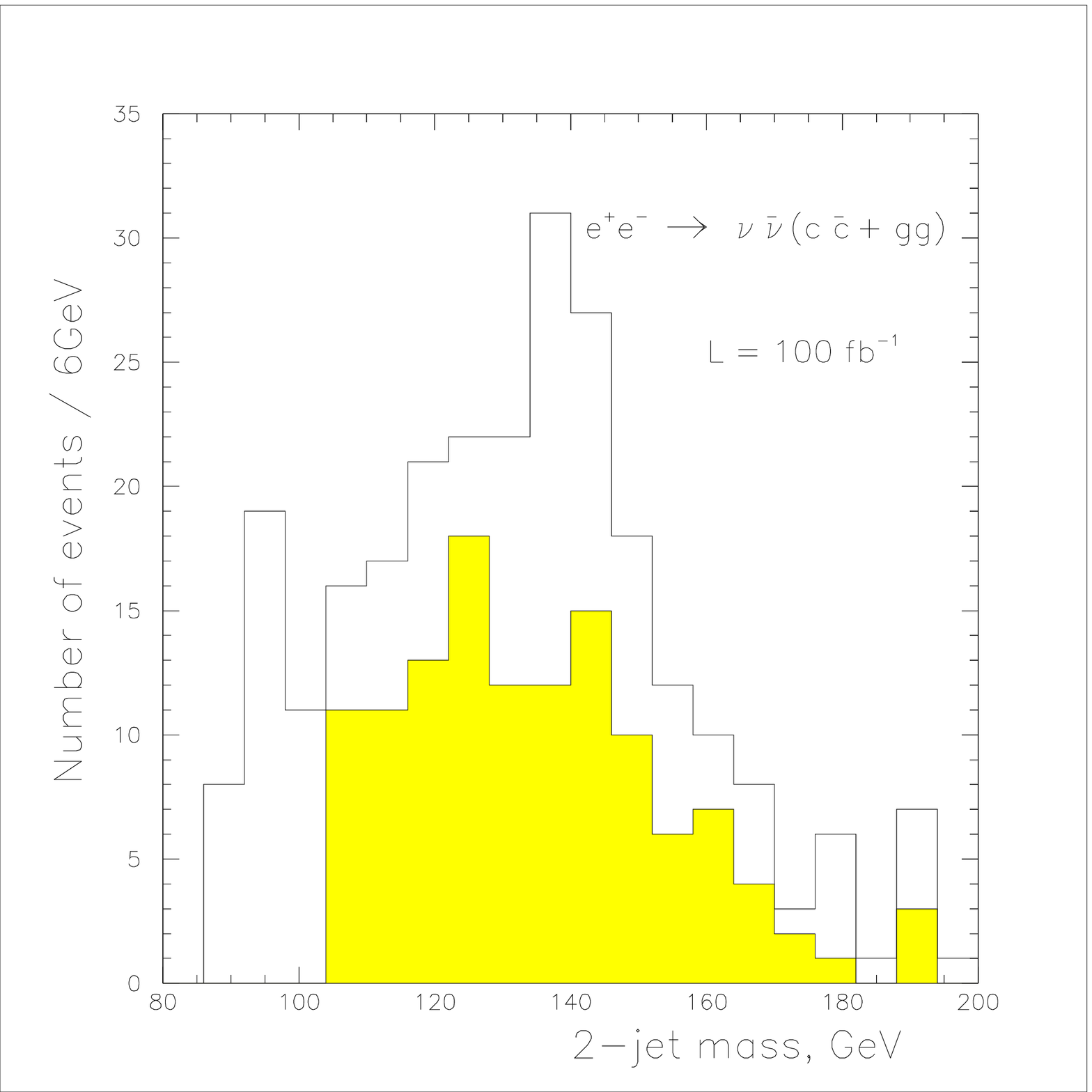}}
\end{center}
\caption[ ]{\sl 2-jet  mass distribution of the reaction
\ee \into \nn(\cc + \glgl) for an integrated luminosity of 100 fb$^{-1}$
at $\sqrt{s} =$ 360 GeV.
The shaded histogram represents the reducible background expected.}
\label{fig:gluon-h}
\end{figure*}

Observation of the $H^o \rightarrow \gamma\gamma$ decay requires large
statistics $(> 200 fb^{-1})$ and a fine-grained electromagnetic
calorimeter with very good energy resolution \cite{Gunion}.

\section{Higgs properties}

If a Higgs candidate is discovered,
 it is imperative to understand its nature. In
the following we assume an integrated luminosity of $100 fb^{-1}$.

Measurements of the Higgs boson mass are best performed by searching for
$Z \rightarrow l^+l^- (l=e/\mu)$ decay products and reconstructing the recoil
mass peak. The mass resolution  depends on
$\sqrt{s}$ and 
the detector performance, in particular on the momentum resolution of
the tracking system. For $M_H = 140$ GeV and $\sqrt{s}$ = 360 GeV,
 we expect
an error for the Higgs mass of $\sim$ 50 MeV for the detector design
of ref.\cite{CDR}.

When the Higgsstrahlung $ZH^o$ rate is significant, the CP-even
component of the $H^o$ dominates. It is possible to cross check this
by studying the Higgs production and the Z decay angular
distributions. Thus, observation of angular distributions as expected
for a pure CP-even state  implies that the
Higgs has spin-0 and that it is not primarily CP-odd \cite{Kramer}.
Further, studying decay angular correlations between the decay
products of the tauons in the reaction
 $e^+e^- \rightarrow ZH^o \rightarrow Z
\tau^+ \tau^-$ provides a democratic probe of the Higgs CP-even and
CP-odd components \cite{Kramer,JFG}. If the $H^o \rightarrow \gamma
\gamma$ decay is visible (or the Higgs is 
produced in $\gamma \gamma$ collisions)
the Higgs must be a scalar and has a CP-even component. Whether a
CP-odd component also exists can be studied by comparing Higgs
production rates in $\gamma \gamma$ collisions with different photon
polarizations \cite{JFG2}.

The determination of the branching fraction of the
 Higgs into a final state X requires to
compute $BF(H \rightarrow X) = [\sigma(ZH) \cdot BF (H \rightarrow
X)] / \sigma (ZH)$, where $\sigma (ZH)$ is the inclusive Higgs
cross section, see Fig. 3. Its error is expected to be $\pm 5\%$
\cite{our}, while the numerator $\sigma (ZH)\cdot BF (H \rightarrow
X)$ and its precision can be obtained from the  X invariant mass
distribution.
\begin{figure*}[h!t]
\begin{center}
\mbox{\epsfxsize=17cm\epsfysize=17.5cm\epsffile{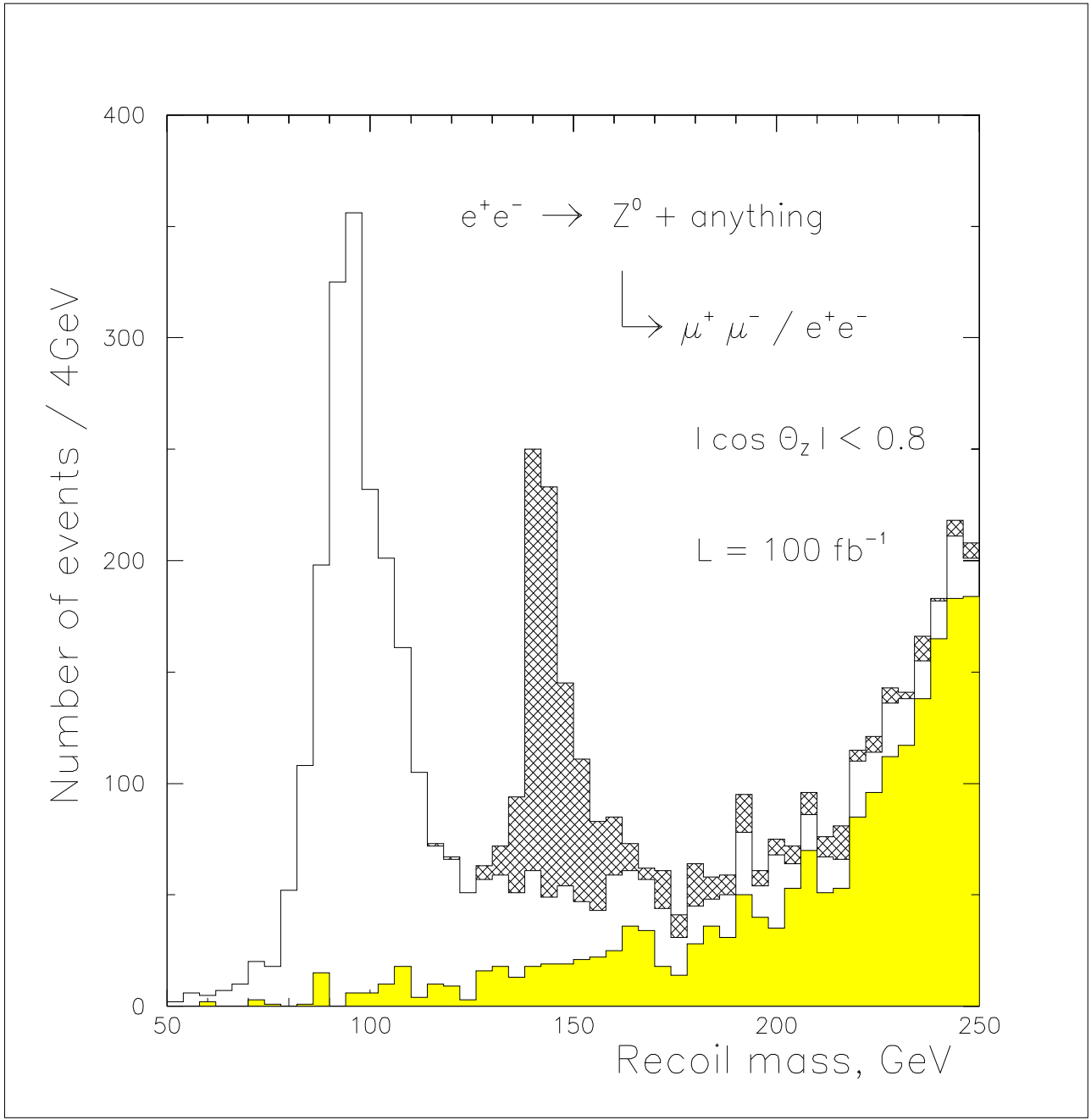}}
\end{center}
\caption[ ]{\sl Inclusive recoil mass distribution in the process
\ee \into \znull(\into \mm/\ee) + anything for an integrated
luminosity of 100 fb$^{-1}$ and $|cos \Theta_Z| <$ 0.8
at $\sqrt{s} =$ 360 GeV.
The shaded histogram represents the reducible background expected,
 whereas the Higgs contribution is shown cross-hatched.
The irreducible background for the \hnull \into WW channel has been
neglected.}
\label{fig:inlusive}
\end{figure*}

Typical statistical branching fraction uncertainties expected for a
140 GeV SM Higgs boson are \cite{our} \\
\\

\begin{tabular}{|c|c|c|c|}\hline \par
$BF(H\rightarrow b \bar{b})$ & $BF(H\rightarrow \tau^+\tau^-)$ & 
$BF(H\rightarrow WW^*)$ & $BF(H\rightarrow c \bar{c} + gg)$ \\ \hline
$\pm 6.1 \%$ & $\pm 22 \%$ & $\pm 22 \%$ & $\pm 28 \%$  \\ \hline
\end{tabular} \\

\vspace{1.0cm}
As can be seen, many of the Higgs decay modes are accessible to
experiment and the measurements allow a
discrimination between the SM-like and e.g. SUSY-like Higgs
bosons. However, more fundamental than branching fractions are the
Higgs partical widths. If $M_H \gess 140$ GeV, $\Gamma(H \rightarrow WW)$
is measurable from the WW fusion cross section, whereas for $M_H \less 
130$ GeV a $\gamma \gamma$ collider allows to measure $\Gamma(H
\rightarrow \gamma \gamma)$.   
Combined with the corresponding branching fraction (from LHC or NLC),
$\Gamma_{tot}^{Higgs}$ is calculable, which in turn allows the
determination of the remaining  partial widths. In addition, the
$H^o \rightarrow ZZ$ coupling-squared is obtained from 
$\sigma (ZH)$ with an error comparable to that of the inclusive
cross section.

\section{MSSM Higgs bosons: $h^o, H^o, A^o, H^{\pm}$}

In the minimal supersymmetric scenario five Higgs particles are
expected, two CP-even states $h^o, H^o$ (with $M_{h^o} < M_{H^o})$, one
CP-odd state $A^o$ and two charged states $H^{\pm}$. An upper bound on
the mass of the $h^o$, including radiative corrections \cite{bound},
is predicted to be $M_{h^o} \less$ 130 GeV.
If $M_{h^o}$
exceeds 130 GeV, the MSSM is ruled out.

In the limit of large $M_{A^o}( \gg M_Z)$ the properties of the $h^o$
are very close to those of the SM Higgs and a discrimination between
the two states is only possible if precise measurements show $h^o
\neq H^o_{SM}$ or further Higgs states are discovered.

Besides Higgsstrahlung and fusion production mechanisms for $h^o$ and
$H^o$, Higgs pair production $e^+e^- \rightarrow h^o A^o,\, H^oA^o,\,
H^+H^-$ is possible, with the relations $\sigma (e^+e^- \rightarrow
h^oZ, h^o\nu\bar{\nu}, H^oA^o) \propto sin^2 (\beta - \alpha)\, (I)$ and
$\sigma (e^+e^- \rightarrow
H^oZ, H^o\nu \bar{\nu}, h^oA^o) \\ \propto
 cos^2 (\beta - \alpha)\, (II)$
$( \alpha, \beta$ are mixing angles). One process in each line (I),
(II) always has a substantial rate and the sum of the $h^o$ or $ H^o$
cross sections is constant and equals  $\sigma (H^o_{SM})$ over the
whole $(M_A, \tan{\beta})$-plane. Therefore, a $h^o$ or $ H^o$ cannot
escape detection if it is kinematically accesible.

Concerning the branching fractions it is usually assumed that the
masses of the supersymmetric particles are so heavy that SUSY-Higgses
cannot decay into s-particles; only decays into standard particles are
possible. Whatever the masses of the Higgs particles and tan$\beta$
values are, we expect i) b-quark decays of $h^o, H^o$ and $A^o$ to occur in
large regions of the parameter space, underlining the importance of
excellent b-tagging, ii) vector boson decays are never significant for
the $A^o$ while strong restrictions exist for the $H^o$ and iii) if 
$M_{H^{\pm}} \gess$ 180 GeV, the $H^{\pm}\rightarrow tb$ decay rate is large
leading in most cases via $ e^+e^- \rightarrow H^+H^-  \rightarrow 
W \,b \bar{b}\, W \,b \bar{b}$ to 8-jet final states involving four b-quark
jets, a very challenging task to recognize and
analyze these events. An example for MSSM Higgs boson detection is
shown in Fig.4 for the process $ e^+e^- \rightarrow H^oA^o \rightarrow
b \bar{b} b \bar{b}$ at $\sqrt{s} =$ 800 GeV \cite{CDR}. 
\begin{figure*}[h!t]
\begin{center}
\mbox{\epsfxsize=17cm\epsfysize=17.5cm\epsffile{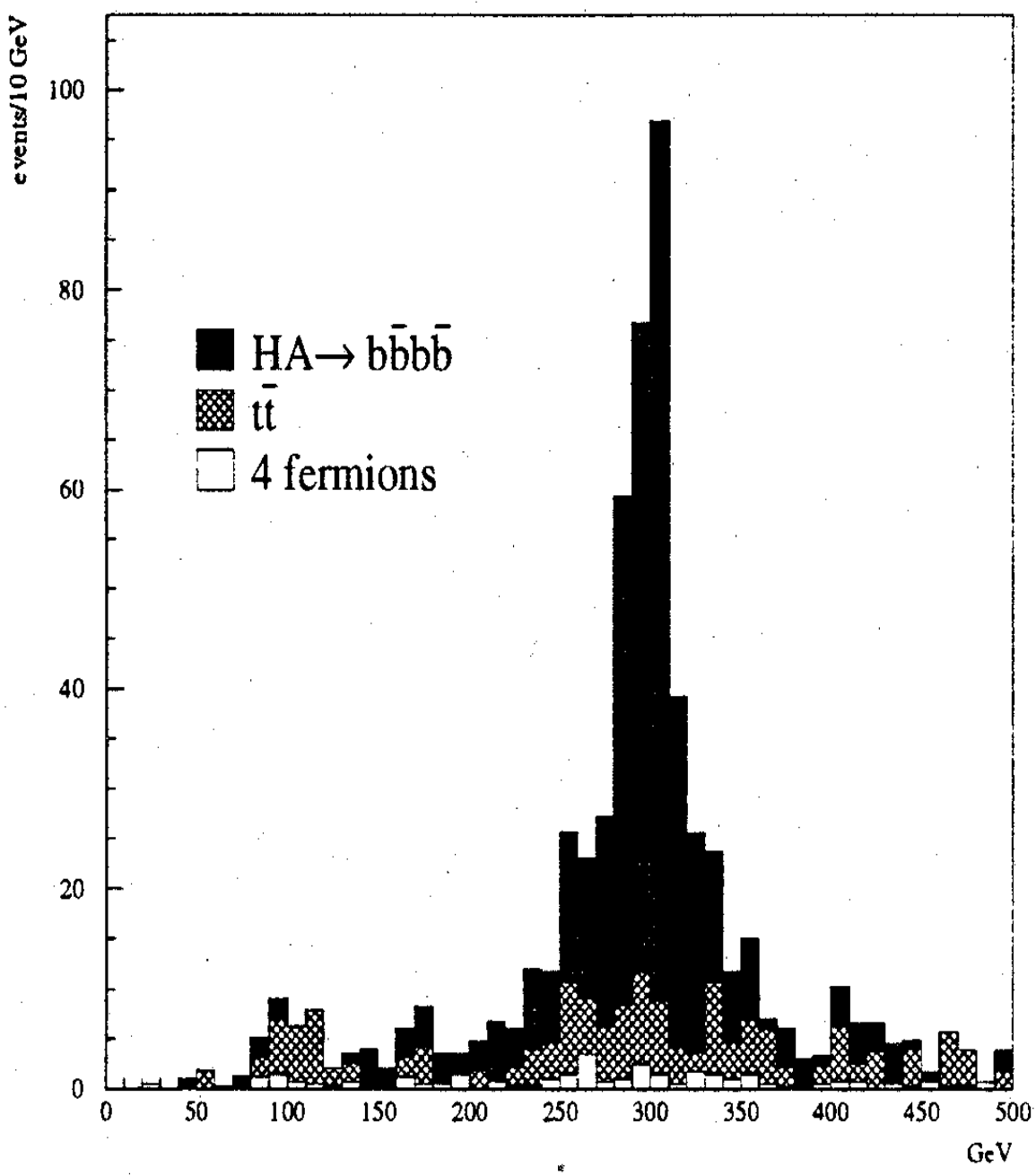}}
\end{center}
\caption[ ]{\sl Distribution of the di-jet invariant mass for the
jet pair combination with better compatibility with the
reaction $ e^+e^- \rightarrow H^oA^o \rightarrow b \bar{b} b \bar{b}$
at $\sqrt{s} =$ 800 GeV.}
\label{fig:eeHA}
\end{figure*}

If however
s-particle decays of Higgses are possible, the decays end (in many
SUSY models) with the lightest 'neutrino-like' supersymmetrie
particle. Hence, event signatures are characterized by missing
$p_{\bot} / E_{\bot}$ + leptons + jets which warrant further
experimental simulations.

\section{Conclusions}

An  $ e^+e^-$ collider focusing on  $ e^+e^- \rightarrow ZH^o$
production in combination with a $\gamma \gamma$ collider allows to
fully explore the properties of an intermediate SM Higgs boson in the
shortest time.

Within the MSSM, one light CP-even Higgs boson must be found or the
MSSM is ruled out. A 500 GeV collider offers very good prospects to
discover all five MSSM Higgs particles, $h^o, H^o, A^o, H^{\pm}$, if 
$M_A \less $ 220 GeV.
If these particles are discoverd large statistics experiments are needed
to measure their parameters.


\end{document}